\documentclass[
   aps,
]{revtex4-2}
\usepackage{graphicx}
\usepackage{tikz}
\usetikzlibrary{calc}
\usepackage{url}            % simple URL typesetting
\usepackage{booktabs}       % professional-quality tables
\usepackage{amsfonts}       % blackboard math symbols
\usepackage{amssymb}
\usepackage{nicefrac}       % compact symbols for 1/2, etc.
\usepackage{microtype}      % microtypography
\usepackage{physics}
\usepackage{bm}
\usepackage[colorlinks]{hyperref}
\hypersetup{citecolor=blue}
\makeatletter
\makeatletter
\providecommand \@ifxundefined [1]{%
 \@ifx{#1\undefined}
}%
\providecommand \@ifnum [1]{%
 \ifnum #1\expandafter \@firstoftwo
 \else \expandafter \@secondoftwo
 \fi
}%
\providecommand \@ifx [1]{%
 \ifx #1\expandafter \@firstoftwo
 \else \expandafter \@secondoftwo
 \fi
}%
\providecommand \bibnamefont  [1]{#1}%
\providecommand \bibfnamefont [1]{#1}%
\providecommand \citenamefont [1]{#1}%
\providecommand \href@noop [0]{\@secondoftwo}%
\providecommand \href [0]{\begingroup \@sanitize@url \@href}%
\providecommand \@href[1]{\@@startlink{#1}\@@href}%
\providecommand \@@href[1]{\endgroup#1\@@endlink}%
\providecommand \@sanitize@url [0]{\catcode `\\12\catcode `\$12\catcode
  `\&12\catcode `\#12\catcode `\^12\catcode `\_12\catcode `\%12\relax}%
\providecommand \@@startlink[1]{}%
\providecommand \@@endlink[0]{}%
\providecommand \url  [0]{\begingroup\@sanitize@url \@url }%
\providecommand \@url [1]{\endgroup\@href {#1}{\urlprefix }}%
\providecommand \urlprefix  [0]{URL }%
\providecommand \Eprint [0]{\href }%
\providecommand \doibase [0]{https://doi.org/}%
\providecommand \selectlanguage [0]{\@gobble}%
\providecommand \bibinfo  [0]{\@secondoftwo}%
\providecommand \bibfield  [0]{\@secondoftwo}%
\providecommand \BibitemOpen [0]{}%
\providecommand \bibitemStop [0]{}%
\providecommand \bibitemNoStop [0]{.\EOS\space}%
\providecommand \EOS [0]{\spacefactor3000\relax}%
\providecommand \BibitemShut  [1]{\csname bibitem#1\endcsname}%
\let\auto@bib@innerbib\@empty\makeatother

\def\mysplit{}

\makeatother
\def\nc{\linebreak}

\begin{document}

%\preprint{}

\title{DNA Torsion-based Model of Cell Fate Phase Transitions}

\author{\textbf{Ng Shyh-Chang}\thanks{Corresponding Author: \href{mailto:huangsq@ioz.ac.cn}{huangsq@ioz.ac.cn}}}
\email{huangsq@ioz.ac.cn}
\affiliation{State Key Laboratory of Stem Cell and Reproductive Biology, \nc Institute of Zoology, Chinese Academy of Sciences,\nc 1 Beichen West Road, Chaoyang District, Beijing 100101, China,\nc Institute of Stem Cell and Regeneration,\nc Chinese Academy of Sciences, Beijing, China,\nc
University of Chinese Academy of Sciences, \nc Beijing, China}
  %% examples of more authors
\author{\textbf{Liaofu Luo}\thanks{Corresponding Author: \href{mailto:lolfcm@imu.edu.cn}{lolfcm@imu.edu.cn}}}
\email{lolfcm@imu.edu.cn}
\affiliation{Faculty of Physical Science and Technology,\nc Inner Mongolia University,  Hohhot 010021, China,\nc
School of Life Science and Technology, \nc Inner Mongolia University of Science and Technology,\nc Baotou, 014010, China}



%\thanks{}


\date{\today}

\begin{abstract}
All stem cell fate transitions, including the metabolic reprogramming of stem cells and the somatic reprogramming of fibroblasts into pluripotent stem cells, can be understood from a unified theoretical model of cell fates. Each cell fate transition can be regarded as a phase transition in DNA supercoiling. However, there has been a dearth of quantitative biophysical models to explain and predict the behaviors of these phase transitions. The generalized Ising model is proposed to define such phase transitions. The model predicts that, apart from temperature-induced phase transitions, there exists DNA torsion frequency-induced phase transitions. Major transitions in epigenetic states, from stem cell activation to differentiation and reprogramming, can be explained by such torsion frequency-induced phase transitions, with important implications for regenerative medicine and medical diagnostics in the future.
\end{abstract}


\maketitle %\maketitle must follow title, authors, abstract


\section{Introduction: DNA torsion as the main variable in stem cell fate decisions}
\subsection{Stem cell fate changes}
Cell biology exploded after Galileo Galilei turned his telescope inward to examine the microscopic world, and after Robert Hooke used his microscope to observe plant and animal tissues for the first time, whereupon he described the existence of `cells'. Since then, $400$ years of biology research have revealed that cells are the basic units of life,
both as free-living single cells and as building blocks within complex multicellular organisms. Developmental biology has addressed many of the questions surrounding how single cells are organized to form multicellular tissues and organisms. By the $21^{st}$ century, it has become clear that the fundamental principles which determine how single cells
with equivalent genomes differentiate into the cornucopia of cell-types in an organism, must lie in the governing dynamics for chromatin epigenetics, gene expression and cell fate transitions in developmental hierarchies\cite{Daley2015}.

The concept that a stem cell at the top of a hierarchy can differentiate into a variety of lineages, owes much to the pioneering ideas and experiments on haematopoietic stem cells by Till and McCulloch\cite{McCulloh1960,Till1961}. Stem cells are characterized by their capacities for long-term self-renewal and multipotent differentiation. By defining the stem cell for blood formation, they also provided an archetype for other developmental systems, including the skin, skeletal muscle, gut, sperm and the early embryo, from which embryonic or pluripotent stem cells are derived. In adult tissues that harbor regenerative potential, stem cells will either undergo a series of fate transitions from activation to proliferation and differentiation during normal development, or switch to senescence, cell death, or cancerous transformation during aging. These fate transitions define the development and aging of every organism on Earth.

Some of the most exciting experiments in developmental biology in the last century include the reprogramming of somatic cells. By using the techniques of somatic cell nuclear transfer (SCNT) or transgenic factor overexpression, biologists were able to reprogram a somatic cell's differentiated fate back to that of a pluripotent stem cell\cite{Robinton2012}. Reprogramming ushered in a new era for human disease modeling and cell-based therapies. It has revolutionized the field of stem cell biology and regenerative medicine, and consequently our notion of the underlying plasticity in chromatin. 

\subsection{Chromatin conformation state and DNA supercoiling}

All eukaryotic cells package their genomes in the form of chromatin, while prokaryotic bacteria package their genomes with similar nucleoid proteins\cite{GasparMaia2010}. Thus genomic DNA is highly compacted in both eukaryotic and prokaryotic cells. DNA compaction state determines its accessibility for transcription, and hence the heterogeneous gene transcriptional states amongst cell populations, despite possessing the same genome. Chromatin consists primarily of DNA and histone proteins. The fundamental unit of chromatin is the nucleosome, which is made of 146 bp of DNA wrapped in supercoiled helical turns around a histone octamer. Each histone's N-terminal tail can undergo covalent modifications which, in turn, control chromatin compaction, eukaryotic gene expression, and play a major role in epigenetic information transfer. For example, histone acetylation is known to locally promote open chromatin conformations and transcription factor binding to activate local gene expression. Physiochemically, the highly basic histone N-terminal tails attractively interact with DNA to facilitate chromatin compaction. Acetylation of the histone N-terminal lysine side-chain removes a positive
charge and thus weakens such electrostatic attractions, resulting in open chromatin\cite{Grunstein1997,Cheung2000}. Most histone modifications depend on cellular metabolism. Metabolites like acetyl-CoA, propionyl-CoA, lactyl-CoA, succinyl-CoA, ATP, ADP-ribose, S-adenosyl-methionine, etc, regulate histone acetylation, propionylation, lactylation, succinylation, phosphorylation, ADP-ribosylation, and methylation, all of which play important roles in the regulation of gene expression and chromatin conformation\cite{ShyhChang2013,Ryall2015,Zhang2018}. 

Because chromatin plasticity is critical for regulating gene expression and thus cell fate transitions, chromatin dynamics have been widely investigated in recent years\cite{Liu2017,Yang2014}. To better understand the fundamental laws that govern dynamic changes in chromatin conformation, we need to develop a deeper understanding of biological macromolecule dynamics, and especially the conformational dynamics of macromolecular DNA\cite{Luo2014,Luo1987,Luo2016}.

The DNA double helix structure is well-suited for its role as a repository of genetic information. After sequencing most major organisms' genomic information, a large portion of the post-genomic effort can be framed as an effort to understand how information-containing DNA is regulated to manage its information transfer to RNA\cite{Frazer2012}. Although the various details of histone modifications and transcription factors are absolutely important, they have also long overshadowed the general principle that these regulatory mechanisms all essentially revolve around regulating chromatin conformation and hence DNA accessibility (i.e. DNA supercoiling conformation). 

DNA supercoiling is the most ubiquitous conformational feature of all eukaryotic and prokaryotic genomes. The supercoiling of DNA around histones in a left-handed direction generates about one negative Wr (writhe) per nucleosome\cite{Nikitina2017}. Local DNA superhelicity not only plays a role in local chromatin conformation, but also in the
stabilization of B-DNA or Z-DNA helical structures to facilitate DNA-protein interactions, especially transcription by RNA polymerases and replication by DNA polymerases\cite{Ravichandran2019}. DNA bond-stretching, bond-bending and torsion angles form a complete set of microscopic variables that define DNA structures. Amongst these microscopic variables, DNA torsion angles are the main determinants that directly affect DNA supercoiling, which affect the accessibility of DNA to polymerase or transcription factor binding\cite{Nikitina2017,Ravichandran2019}. Thus, gene expression changes and cell fate transitions are the general results of changes in DNA torsion angles.

\subsection{DNA torsion energy and the Hamiltonian}

We arrived at this conclusion based on existing biological knowledge and intuitive biochemical reasoning. Yet, to analyze the state changes in cell populations, it is imperative to introduce the theory of phase transitions and self-organization in physics. We shall begin by describing each cell as a dynamical system defined with the formalism of Hamiltonian mechanics, whereby a system is described by a set of canonical coordinates $\bm{r = (q, p)}$ in phase space.
The time evolution of the dynamical system is uniquely defined by: 

\[
\dv{\bm{p}}{t}=-\pdv{\mathcal{H}}{\bm{q}}\quad,\quad\dv{\bm{q}}{t}=+\pdv{\mathcal{H}}{\bm{p}}
\]
where $\mathcal{H} = \mathcal{H}(\bm{q}, \bm{p}, t)$ is the Hamiltonian function, representing the total energy of the system. Hamiltonians find applications in all areas of physics, from celestial mechanics to quantum mechanics, and especially in complex dynamical systems. For molecular systems, one should start from the principle of quantum mechanics and the momentum $\bm{p}$
in Hamiltonian should be replaced by an operator $\pdv{}{\bm{q}}$.  In this case the Hamiltonian is also an operator. With increasing degrees of freedom, a Hamiltonian system's time evolution becomes more complicated and often chaotic\cite{Latora1999}. Systems with many (sometimes infinite) degrees of freedom or variables are generally hard to solve or compute exactly. Statistical mechanics methods are generally introduced to solve such many-body problems. 

One classic example is the Ising model\cite{Torquato2011}. The Ising model was first used to predict how ferromagnetism arises through a phase transition in a system of particles, each particle with its own up or down magnetic spin. The term `phase transition' is most commonly used to describe abrupt transitions between different states of matter, e.g. solid, liquid, and gas. Phase transitions occur when the free energy of a system shows discontinuity with respect to some variable, e.g. temperature or pressure. Phase transitions generally stem from the interactions of a large number of particles in a complex system, and does not appear in systems that are too small. 

Amongst the large variety of variables in the high-dimensional space of a complex dynamical system, abrupt transitions only manifest in the `order parameters'. An order parameter shows the degree of order across the boundaries in a phase transition system; it normally ranges between zero in one phase, and nonzero in the other phase, separated by the critical point\cite{Torquato2011}. An example of an order parameter is the net magnetization in a ferromagnetic system undergoing a phase transition. For liquid/gas transitions, the order parameter is the level of densities. This order parameter concept, originally introduced in the Ginzburg--Landau theory for phase transitions in thermodynamics, was generalized by Haken to the ``enslaving principle'', which states that the dynamics of fast-relaxing variables is completely determined by the slow-relaxing dynamics of only a few 'order parameters'\cite{Haken1987}. 

We shall assume DNA is the major macromolecular chain that determines a cell's (gene expression) state. For each monomer or nucleotide of DNA, the bond lengths, bond angles, torsion angles $\{\theta\}$, and the coordinates of electrons/molecules bound to the DNA, define a complete set of microscopic variables to describe its Hamiltonian system. Torsion vibration energy is $0.003$--$0.03\; eV$, the lowest in all forms of biological energies, even lower than the average thermal energy per atom at room temperature ($0.04\; eV$ at $25\;{}^{\circ}C$). Thus, torsion angles are easily changed even at physiological temperature, and represent slow-relaxing or unstable variables. Following Haken's enslaving principle, torsion angles would represent the `order parameters' in the DNA molecular system. Moreover, the torsion motion has two other important peculiarities. First, our earlier work had already proven that a macromolecular chain, including DNA, would manifest a rapid increase in Shannon information quantity at room temperature as its oscillator frequency decreases below $10^{13}\; Hz$, through a Bose-Einstein condensation of phonons\cite{Luo1987}. The DNA torsion vibration frequency is exactly in the range below $10^{13}\; Hz$. Therefore, the torsion vibration conveys the largest information quantity, as compared to bond bending and stretching, and it may play an important role in the transmission of genetic information and genetic noise within cells\cite{Tkaik2008}. Second, unlike stretching and bending, the torsion potential generally has several minima with respect to angle coordinates that correspond to several stable conformations. In other words, a cell's state is determined mainly by phase transitions between minima in its DNA torsion energy state, not other variables, which is a more quantifiable form of the same conclusion we arrived at with intuitive biochemical reasoning above\cite{Luo1987,Luo2016}.

Based on this argument, we will propose a model on the mechanisms of stem cell differentiation and cell fate transitions in general, based on phase transitions in DNA torsion.

\section{Methods}

The Hamiltonian of the DNA molecular system can be expressed as

\begin{equation}\label{eq:1}
\mathcal{H}=\mathcal{H}_S\left(\theta,\pdv{}{\theta}\right)+\mathcal{H}_F\left(x,\pdv{}{x};\theta\right)
\end{equation}
where  $\mathcal{H}_S$ is the
slow-relaxing variable (denoted as $\theta$) Hamiltonian, including the torsion angles of each nucleotide, $\mathcal{H}_F$  is the fast-relaxing
variable Hamiltonian (denoted as $x$) including the bond stretching / bending ,the electronic variables, etc. The
stationary Schrodinger equation 

\begin{equation}\label{eq:2}
\mathcal{H}M(\theta,x)=EM(\theta,x)
\end{equation}
can be solved under the adiabatic approximation,

\begin{equation}\label{eq:3}
M(\theta,x)=\psi(\theta)\phi(x,\theta)
\end{equation}
and these two factors satisfy

\begin{equation}\label{eq:4}
  \mathcal{H}_F\left(x,\pdv{}{x};\theta\right)\phi_{\alpha}(x,\theta)=\epsilon^{\alpha}(\theta)\phi_{\alpha}(x,\theta)
\end{equation}
\begin{equation}\label{eq:5}
  \left\{\mathcal{H}_S\left(\theta,\pdv{}{\theta}\right)+\epsilon^{\alpha}(\theta)\right\}\psi_{kn\alpha}(\theta)=E_{kn\alpha}\psi_{kn\alpha}(\theta)
\end{equation}
respectively\cite{Luo2014}. Here $\alpha$ denotes the quantum state of fast-relaxing variables, and $(k, n)$ refer to the quantum
numbers of torsional conformation and torsional vibration of the DNA molecular system. For a DNA molecular chain of
nucleotides, Eq\eqref{eq:5} can be rewritten into

\begin{subequations}
  \begin{equation}
  \label{eq:6a}
  \begin{split}
  \sum\left(-\frac{\hslash^2}{2I_j}\pdv[2]{}{\theta_j}+U_{tor}(\theta_1,\ldots,\theta_s)\right)\psi(\theta_1,\ldots,\theta_s)\mysplit
  =K_{kn\alpha}\psi_{kn\alpha}(\theta_1,\ldots,\theta_s)
  \end{split}
  \end{equation}
  \begin{equation}\label{eq:6b}
    U_{tor}(\theta_1,\theta_2,\ldots,\theta_s)=\sum_j U_{tor}^{(j)}(\theta_j)+\sum_j U_{tor}^{(j,j+1)}(\theta_j,\theta_{j+1})
    \end{equation}
\end{subequations}

Note that here the potential $U_{tor}(\theta_1,\ldots,\theta_s)$  is dependent on the
fast-relaxing variable quantum number $\alpha$ through the term $\epsilon^{\alpha}(\theta)$  as indicated in Eq\eqref{eq:5}.
Eq\eqref{eq:6b} shows that the torsion potential $U_{tor}(\theta_1,\ldots,\theta_s)$ includes two parts, the
term $U_{tor}^{(j)}(\theta_j)$  of a single
nucleotide within the chain and the interaction $U_{tor}^{(j,j+1)}(\theta_j,\theta_{j+1})$  between neighboring
nucleotides. As the interaction is switched off, the solution of Eq\eqref{eq:6a} can be expressed as the product of each single
nucleotide's wave functions, $\psi_{kn\alpha}(\theta_1,\ldots,\theta_s)=\prod_j \psi_{k_j n_j \alpha_j} (\theta_j)$. The general solution
of Eq\eqref{eq:6a} is the linear combination of $\psi_{kn\alpha}(\theta_1,\ldots,\theta_s)$.  The quantum number $k_j$ is referred to the conformation state of the $j$-th nucleotide and $n_j${}-its vibration state. 

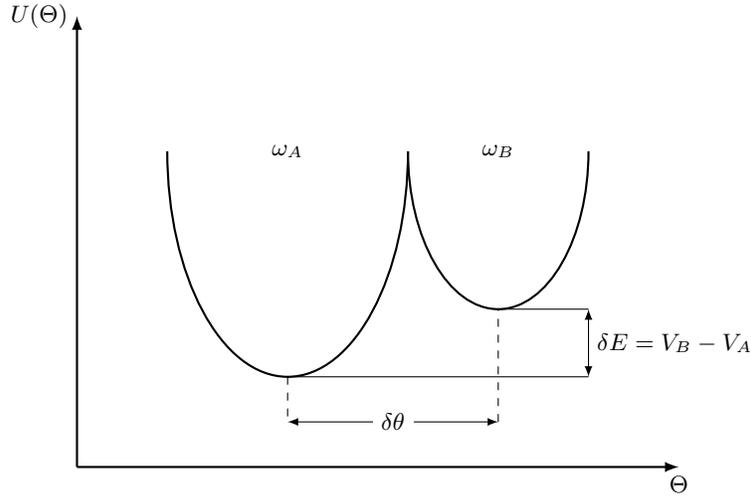
\begin{figure}
  \centering
  \def\Width{8}%
  \def\Height{6}%
  \begin{tikzpicture}
    \draw[thick,-latex] (0,0)--(\Width,0)node[below]{$\Theta$};
    \draw[thick,-latex] (0,0)--(0,\Height)node[left]{$U(\Theta)$};
    \draw[thick] ({0.75*\Width/5},{3.5*\Height/5}) arc(180:360:{1*\Width/5} and {2.5*\Height/5})node[midway,shift={(0,{2.5*\Height/5})}]{$\omega_A$};
    \draw[thick] ({2.75*\Width/5},{3.5*\Height/5}) arc(180:360:{0.75*\Width/5} and {1.75*\Height/5})node[midway,shift={(0,{1.75*\Height/5})}]{$\omega_B$};
    \draw[thin,dashed] ({1.75*\Width/5},{\Height/5})coordinate(UA)--({1.75*\Width/5},{\Height/5-\Height/10}) coordinate(LA);
    \draw[thin,dashed] ({3.5*\Width/5},{\Height/5-\Height/10})coordinate(LB)--({3.5*\Width/5},{1.75*\Height/5})coordinate(UB);
    \draw[latex-latex] (LA)--(LB)node[midway,fill=white]{$\delta\theta$};
    \draw (UA)--($(UA)+({2.5*\Width/5},0)$) coordinate(RA);
    \draw (UB)--($(UB)+({0.75*\Width/5},0)$)coordinate(RB);
    \draw[latex-latex](RA)--(RB)node[midway, anchor=west]{$\delta E=V_B-V_A$};
  \end{tikzpicture}
\caption{Torsion potential energy $U(\theta)$ versus torsion angle $\theta$.}\label{fig:1}
\end{figure}

Based on the above formulation, we can study the DNA molecule in detail. Assume the torsion potential $U_{tor}^{(j)}(\theta_j)$  (j=1,\ldots, s) has
two minima $V_A$ and $V_B$ as shown in Figure~\ref{fig:1}. The corresponding vibration frequencies around two minima are denoted as
$\omega_A$ (in left well) and $\omega_B$B (in right well) respectively. We propose that the structural foundation of the
activation/differentiation of stem cells is the existence of pairs of torsion quantum states (torsion ground-state and
torsion excited-state) for each nucleotide within a gene region. That is, we assume the quantum number kj takes two
values, $k_j=A$ or $B$ describing these two states. Under this assumption, the macroscopic epigenetic state of stem cells
could be understood as the combinatorial result of quantum transitions between these two microscopic DNA torsion
states. Of course, apart from DNA torsion, there exists other molecular variables that may influence the
activation/differentiation of stem cells. It includes chemical reactions that result in changes in protein electronic
configurations, small molecule binding interactions, chromatin configuration and other epigenetic factors, etc. All
these variables are either fast-relaxing variables or their influence can be ultimately represented and estimated with
DNA torsion.

\begin{align}
  U(\theta)=V_A+\frac{1}{2}I\omega_A^2(\theta-\theta_A)^2\tag{left}\\
  U(\theta)=V_B+\frac{1}{2}I\omega_B^2(\theta-\theta_B)^2\tag{right}
\end{align}

\section{Results}
\subsection{Statistical mechanics of DNA molecules}
Let us assume that the system of DNA chain of nucleotides is in thermal equilibrium. We will calculate the probabilities
of the DNA chain in two torsion states $A$ and $B$. Denote the partition functions (summation of probabilities) for a
single section as $Z_A$ and $Z_B$ respectively. We have\cite{Luo1987,Luo1985}:

\[
\frac{Z_A}{Z_B}=e^{-\beta (V_A-V_B)}Y_{A/B}\;\;\;\;\;\left(\beta=\frac{1}{k_B T}\right)
\]

\begin{equation}\label{eq:7}
  Y_{A/B}=\frac{e^{\frac{\beta\hslash\omega_B}{2}}-e^{\frac{-\beta\hslash\omega_B}{2}}}{e^{\frac{\beta\hslash\omega_A}{2}}-e^{\frac{-\beta\hslash\omega A}{2}}}
  \end{equation}

$Y_{A/B}$ comes from the summation over vibration states. If the conformation vibration is neglected, then the probability
ratio is simply determined by $V_A-V_B$. Suppose $V_A<V_B$ , then $A$ is the favored conformation since
$Z_A/Z_B>1$. However, the vibrations around the potential minimum are important for the fixation of a definite
conformation. Since $Y_{A/B}>1$ as $\omega_A<\omega_B$ and $Y_{A/B}<1$ as $\omega_A>\omega_B$, the conformation with lower vibration frequency is more favored. When $\omega_B$ is much
smaller than $\omega_A$ , one has $Z_A/Z_B<1$ and the conformation $B$ is the favored one instead of $A$.

The above analysis was made for a single nucleotide of the DNA molecular chain. Next we will discuss the cooperativity
between nucleotides in a DNA molecular chain. The partition function of the DNA molecular chain is:

\begin{subequations}
  \begin{equation}
  \begin{split}
    Z &=\sum_{k_l n_l}\ldots\sum_{k_s n_s} e^{-\beta \sum_i E_{k_i n_i}} e^{-\beta\sum_i U_{k_i n_i k_{i+1} n_{i+1}}}\\
    &=\sum_{k_1=A,B}\sum_{k_2=A,B}\ldots\sum_{k_s=A,B} e^{-\beta\sum_i {E^{\prime}}_{k_i}} e^{-\beta\sum_i {\mathit{U}}_{k_i,k_{i+1}}}
  \end{split}
  \end{equation}
  \begin{equation}
  \begin{split}
    \exp\left(-\beta{E^{\prime}}_{k_i}\right)=\exp\left(-\beta V_{k_i}\right)\left(e^{\frac{1}{2}\beta\hslash{\omega}_{k_i}}-e^{-\frac{1}{2}\beta\hslash\omega_{k_i}}\right) \;\;\;\;\mysplit
    (k_i=A,B)
    \end{split}
  \end{equation}
  \begin{equation}
  \begin{split}
  \exp\left(-\beta{\mathit{U}}_{k_i k_{i+1}}\right)=\langle\exp\left(-\beta {\mathit{U}}_{k_i n_i k_{i+1} n_{i+1}}\right)\rangle\;\;\;\;\mysplit 
  (k_i=A,B;k_{i+1}=A,B)
  \end{split}
  \end{equation}
\end{subequations}

where $\langle\phantom{ }\rangle$  means the average
over vibrational states. Here we introduce the matrix $P_i$ where

\[
\langle k_i |P_i|k_{i+1}\rangle=\exp\left(-\beta {E^{\prime}}_{k_i}\right)\exp\left(-\beta {\mathit{U}}_{k_i k_{i+1}}\right)
\]

Under the periodic boundary conditions one has

\begin{equation}
Z= Tr\left(P_1,\ldots,P_{s}\right)
\end{equation}

If s sections are same, then $P_i=P$

\begin{equation}
  \begin{split}
    P=\begin{pmatrix}
    \exp\left(-\beta {E^{\prime}}_A\right) & \exp\left(-\beta({E^{\prime}}_A+\mathit{U})\right)\\
    \exp\left(-\beta({E^{\prime}}_B+\mathit{U})\right) & \exp\left(-\beta {E^{\prime}}_B\right)
    \end{pmatrix}\equiv\mysplit
    \begin{pmatrix}
    1 & \sigma\\
    \zeta_{\sigma} & \zeta
    \end{pmatrix}\exp\left(-\beta {E^{\prime}}_A\right)
  \end{split}
\end{equation}

($U_{AB} = U_{BA} = U$ is assumed and $U_{AA}$ and $U_{BB}$ are neglected).  One has

\begin{equation}
  Z=Tr\left(P^{S}\right)=\left(\lambda_{max}\right)^{S}
\end{equation}

where  $\lambda_{max}$  is the largest
eigenvalue of matrix $P$. The probabilities of nucleotides in state $A$ (denoted as $O_A$) or $B$ (denoted as $O_B$) are deduced
from 

\[
O_A=-\frac{1}{s\beta}\pdv{\ln{Z}}{E_A}=-\frac{1}{\beta}\pdv{\ln{\lambda_{max}}}{{E_A}^{\prime}}
\]

\begin{equation}
O_B=1-O_A.
\end{equation}

The calculation method given above is the same as the method used to solve the Ising model\cite{Huang1987}. Finally we obtain the
order parameter\cite{Luo1987,Luo1985}

\[
O_A=\frac{1}{2}-\frac{1}{2}\frac{\sinh{\left(\frac{\beta}{2}({E^{\prime}}_A-{E^{\prime}}_B)\right)}}{\sqrt{\sinh^2{\left(\frac{\beta}{2}({E^{\prime}}_A-{E^{\prime}}_B)\right)}+\exp(-2\beta \mathit{U})}}
\]

\begin{equation}\label{eq:13}
O_B=\frac{1}{2}+\frac{1}{2}\frac{\sinh{\left(\frac{\beta}{2}({E^{\prime}}_A-{E^{\prime}}_B)\right)}}{\sqrt{\sinh^2{\left(\frac{\beta}{2}({E^{\prime}}_A-{E^{\prime}}_B)\right)}+\exp(-2\beta \mathit{U})}}
\end{equation}

where

\begin{equation}
{E^{\prime}}_A-{E^{\prime}}_B=V_A-V_B-k_B T \ln{Y_{A/B}}.
\end{equation}

Since the parameters $O_A$ or $O_B$ are decisive factors in DNA structure, they can be regarded as the order parameters of the
system. If the torsion correlation between neighboring nucleotides is strong enough, $U\gg kBT$, then  $\exp(-2\beta \mathit{U})=0$ 

\[
 O_A=1, O_B=0 \text{ as  } {E^{\prime}}_A-{E^{\prime}}_B<0  
 \]

 \begin{equation}
 O_A=0, O_B=1 \text{ as  } {E^{\prime}}_A-{E^{\prime}}_B>0  
 \end{equation}

(the chain is condensed fully in phase $A$ or phase $B$ respectively). So, there exists two phases $A$ and $B$ given by the
symbol of  ${E^{\prime}}_A-{E^{\prime}}_B$.  Of course,
the condensation may never be complete in general since the small term 
$\exp\left(-2\beta \mathit{U}\right)$  in Eq\eqref{eq:13} may only
approach but never equal zero. 

To summarize, for $V_A<V_B$ (Figure \ref{fig:1}), the system would condense into state $A$ as the vibration is switched off.
However, the vibration term $Y_{A/B}$ changes the result as 
$\omega_A \neq \omega_b$. Under $\lvert\frac{\omega_A-\omega_B}{\omega_A}\rvert\ll 1$  from Eq\eqref{eq:7} we have

\begin{equation}\label{eq:16}
k_B T\ln{Y_{A/B}}=\frac{\hslash}{2}\left(\omega_A-\omega_B\right)\mathit{ctnh}\frac{\hslash\omega_A}{2k_B T}
\end{equation}
where the function $\mathit{ctnh}x$ is defined by $\mathit{ctnh}x=\frac{e^x+e^{-x}}{e^x-e^{-x}}$, an odd function
decreasing with $x$ and always larger than $1$ for positive $x$. Eqs\eqref{eq:13} to \eqref{eq:16} constitute our main results on the
cooperative mechanism or phase transition of DNA molecules.

\subsection{Phase transitions and applications in cell fate decisions}
In statistical physics there is a theorem that states: no phase transition exists in a 1D Ising model\cite{Xu2014}. However, from
the above generalized Ising model, we have shown that a phase transition can also occur in the 1D chain, as torsion
vibration is taken into account. In fact, from Eqs\eqref{eq:13} to \eqref{eq:16} the phase transition occurs at ${E^{\prime}}_A={E^{\prime}}_B$, namely

\begin{equation}\label{eq:17}
  V_A-V_B=\frac{\hslash}{2}\left(\omega_A-\omega_B\right)\mathit{ctnh}\frac{\hslash\omega_A}{2k_B T}
\end{equation}

As  $V_A-V_B<\frac{\hslash}{2}(\omega_A-\omega_B)\mathit{ctnh}\frac{\hslash\omega_A}{2k_BT}$, the chain
condenses into A-phase and as  $V_A-V_B>\frac{\hslash}{2}(\omega_A-\omega_B)\mathit{ctnh}\frac{\hslash\omega_A}{2k_B T}$, the chain condenses into B-phase.

This system exhibits two kinds of phase transitions. The first is the temperature-induced phase transition (T-phase transition), occurring at the critical temperature $T_c$

\begin{equation}\label{eq:18}
T_c=\frac{\hslash\omega_A}{2k_B}\frac{1}{\mathit{ctnh}^{-1}\frac{2(V_B-V_A)}{\hslash(\omega_A-\omega_B)}}
\end{equation}

Since $\mathit{ctnh}x\geq 1$ (for positive $x$), the phase transition only exists under the condition 
$V_B-V_A>\frac{\hslash}{2}\left(\omega_A-\omega B\right)>0$ or  $V_A-V_B>\frac{\hslash}{2}\left(\omega_A-\omega_B\right)>0$. For example, as $V_B>V_A$ the chain is condensed in A-phase as $T<T_C$T and in
B-phase as $T>T_c$ for $\omega$ frequencies that satisfy the above conditions (Figure~\ref{fig:2}). The prediction
that there exists a temperature-induced phase-transition provides an experimental checkpoint for our present theory.
Moreover, one may deduce the ratio $\frac{2(V_B-V_A)}{\hslash(\omega_A-\omega_B)}$  from Eq\eqref{eq:18} by using the
measured value of the critical transition temperature $T_c$.

 The second type of phase transition predicted for the DNA molecular chain is the torsion--induced phase transition. One
 can adjust the frequency $\omega_A$ or $\omega_B$ of torsion potential (and/or $V_B-V_A$) to obtain the phase transition.
 For example, as $V_A<V_B$, the chain is condensed in state $A$ as $\omega_A = \omega_B$. However, we can adjust
$\omega-B$ to induce an $\omega$-phase transition. Suppose  $\frac{\hslash\omega_A}{k_B} = 0.1$ to $1$. We
predict that the $\omega $-phase transition from state A to B can be realized through decreasing $\omega_B$ by $\delta\omega =\frac{1}{\hslash}(0.1-0.92)(V_B-V_A)$. In fact, by setting $\delta \omega =\omega_A-\omega_B$, the critical torsion point $(\delta \omega)_c$ at a given temperature $T$ is defined by 

\begin{equation}\label{eq:19}
  (\delta \omega)_c=\frac{2(V_B-V_A)}{\hslash\mathit{ctnh}\frac{\hslash\omega_A}{2k_B T}}.
\end{equation}
which is deduced from Eq\eqref{eq:17} (Figure~\ref{fig:2}). The prediction of $\omega $-phase-transitions provides another experimental
checkpoint for the present theory.

To examine the experimentally verifiable and quantitative relationship between the critical temperature Tc and the
critical torsion point in detail, it would be useful to further simplify Eq\eqref{eq:17}:

 Let  $\frac{\hslash(\omega_A-\omega_B)}{2(V_B-V_A)}=\Delta_{\omega }$ (dimensionless torsion ratio)

  $\frac{\hslash\omega_A}{2k_B}=\alpha$ (constant)

 then  
\begin{equation}\label{eq:20}
 \Delta_{\omega }=\tanh\frac{a}{T}.
\end{equation}

This is a sigmoidal function that would plateau out at extreme values of T. But we should note that for torsion
vibration energies of $0.03$--$0.003\; ev$, the constant a is estimated to be
$\sim\!\!100K$. If we plot the possible distribution of values of the critical torsion ratio $\Delta_{\omega}$ for $T$ in the physiological range of $273$ to $323K$ ($0$ to $50\;{}^{\circ}C$),
we obtain a decreasing curve:

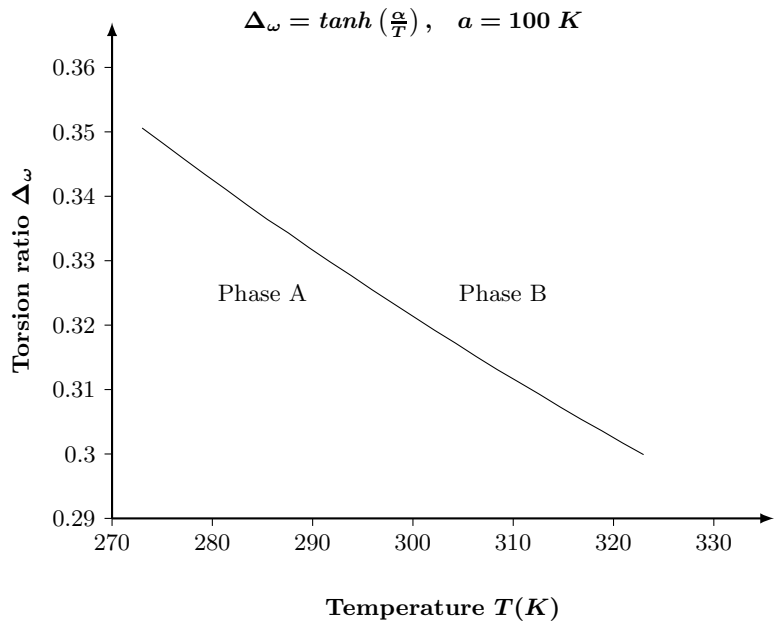
\begin{figure}
  \centering
  \def\Width{8}%
  \def\Height{6}%
  \def\TickSize{0.1}%
  \begin{tikzpicture}
    \draw[thick,-latex](0,0)--({1.1*\Width},0)node[midway,shift={(0,-1.2)}]{\bfseries Temperature $\bm{T (K)}$};
    \draw[thick,-latex](0,0)--(0,{1.1*\Height})node[midway,shift={(-1.2,0)},rotate=90]{\bfseries Torsion ratio $\bm{\Delta_{\omega}}$};
    \foreach \x in {0,...,6}{%
      \pgfmathtruncatemacro\T{270+10*\x}%
      \draw ({(\x*\Width/6)},0)--({(\x*\Width/6)},{-\TickSize})node[below]{$\T$};
    }
    \foreach \y in {0,...,7}{%
      \pgfmathsetmacro\TR{0.29+\y/100}%
      \draw (0,{(\y*\Height/7)})--({-\TickSize},{(\y*\Height/7)})node[left]{$\pgfmathprintnumber[fixed, precision=2]{\TR}$};
    }
    \draw plot[thick,domain={(273)/60*\Width}:{(323)/60*\Width},shift={({-270/60*\Width},0)}] ({\x},{(tanh(((100)/60*\Width)/(\x)))*\Height/7/0.01-0.29*\Height/7/0.01});
    \node at ({.25*\Width},{0.5*\Height}){Phase A};
    \node at ({.65*\Width},{0.5*\Height}){Phase B};
    \node at ({.5*\Width},{1.1*\Height}){$\bm{\Delta_{\omega}=\mathit{tanh}\left(\frac{\alpha}{T}\right),\;\;\;a=100\;K}$};
  \end{tikzpicture}
\caption{Phase diagram for any DNA region, with the order parameters defined as temperature $T$ and the torsion ratio $\Delta_{\omega}$. Based on Eq\protect\eqref{eq:20}, a decreasing curve separates phases $A$ and $B$, indicating that as the temperature $T$ increases, the critical torsion ratio $\Delta_{\omega}$ required for DNA to undergo a phase transition becomes smaller. The torsion ratio is defined by $\Delta_{\omega}=\frac{\hslash(\omega_A-\omega_B)}{2(V_B-V_A)}$, where $\hslash$ is the reduced Planck constant, ($\omega_A-\omega_B$) is the change in torsion
frequency, and ($V_B-V_A$) is the difference in torsion potential energy.}\label{fig:2}
\end{figure}

Moreover we note that  $\dv{\Delta_{\omega}}{T}=\frac{-\alpha}{T^2}\mathit{sech}^2\frac{\alpha}{T}$ .  

Suppose there exists a molecule in the microenvironment that can induce DNA torsion changes, and its relationship with
torsion can be represented linearly as

$\Delta_{\omega }=\Delta_o+bC$  where $\Delta_o$ is the initial torsion ratio, $C$ is the concentration
of the molecule, and $b$ is the linear coefficient

then

\begin{equation}\label{eq:21}
  \dv{C}{T}=\frac{-\alpha}{bT^2}\mathit{sech}^2\frac{\alpha}{T}.
\end{equation}

With Eq\eqref{eq:20} and Eq\eqref{eq:21}, one should be able to experimentally verify if the critical molecular concentration $C$ vs.
critical temperature $T$ curve follows the quantitative relationship predicted by this theory.

Suppose we define the phase index ($PI$) as

\begin{equation}\label{eq:22}
\mathit{PI}=1-\frac{\hslash(\omega_B-\omega_A)}{2(V_A-V_B)}\mathit{ctnh}\frac{\hslash\omega_A}{2k_BT }.
\end{equation}

In the vicinity of the critical point, by inserting Eq\eqref{eq:18} into \eqref{eq:19} we obtain

\begin{equation}\label{eq:23}
\mathit{PI}=1-\frac{\mathit{ctnh}\frac{\hslash\omega_A}{2k_BT }}{\mathit{ctnh}\frac{\hslash\omega_A}{2k_BT_C}}.
\end{equation}

$PI$ is therefore an observable parameter that indicates where in phase space the system resides, relative to the critical
phase transition point. When $T=T_C$ ($PI=0$), the system is at the critical phase transition point. When $T<T_C$ ($PI<0$), the system is condensed in phase $A$. When $T<T_C$ ($PI>0$), the system is condensed
in phase $B$. 

Abrupt cell fate changes during somatic reprogramming, directed differentiation or activation of stem cells (all typically cultured at $\sim\!\!37\;{}^{\circ}C$), provide a variety of vignettes that suggest the existence of cellular phase transitions\cite{Xu2014}. In fact, there are many molecules that can induce DNA torsion changes to trigger $\omega$-phase transitions. In particular the metabolic reprogramming of stem cells, including pluripotent stem cells and muscle stem cells, suggest that non-lineage-specific factors could trigger such $\omega$-phase transitions. For example, histones are modified by histone acetyltransferases (HAT) and deacetylases (HDAC) to direct chromatin conformation and regulate local supercoiling of DNA\cite{Grunstein1997,Cheung2000}. Here, the metabolite-based regulation of HATs and HDACs by acetyl-CoA, NAD+, and short-chain
fatty acids can directly induce changes in DNA torsion potential to trigger $\omega $-phase transitions in cell fate\cite{Morales2000}. Histone methylation and its regulation by histone methyltransferases (HMT) and JmjC-domain-containing histone demethylases (JHDM) are alternative mechanisms to regulate chromatin conformation and local DNA supercoiling. Similar to histone acetylation, metabolite-based regulation of HMTs and JHDMs via S-adenosyl-methionine, ${\alpha}$-ketoglutarate, ascorbate and $Fe^{2+}$ can also induce changes in DNA torsion potential to trigger cell fate transitions\cite{Brumbaugh2019}. It is also well-known that direct genetic mutation of histones or histone modifying enzymes to alter the chromatin conformation and DNA supercoiling can induce drastic changes in stem cell fate decisions and organismal development\cite{Buschbeck2017}. Some anthracycline drugs such as doxorubicin or idarubicin can directly intercalate between nucleotide base pairs in DNA and alter DNA torsion potential to induce mitotic arrest or differentiation in cancer cells.\cite{Yang2014}

From a general theoretical point of view, all molecule-induced changes in the torsion parameters $\omega_A,\omega_B,V_A,V_B$ can cause a phase transition. These changes can occur through the change of torsion potential $U_{tor}(\theta_1,\ldots,\theta_S)$. In fact, all the metabolite- or genetic- or drug-induced torsion changes discussed above could be ascribed to the $\epsilon^{\alpha}(\theta)$  term for electron motion (between histones and DNA), and this term was already included as a part of the potential term $U_{tor}(\theta_1\ldots,\theta_S)$. 

NB: Molecular binding is an important factor in stem cell activation and differentiation. The binding or unbinding of (small) molecules to DNA creates an additional term in the Hamiltonian Eqs\eqref{eq:4} and \eqref{eq:5}. The partition function of the DNA molecular chain will be changed from Eq\eqref{eq:11-1} to 

\begin{subequations}
  \begin{equation}\label{eq:11-1}
    Z=\mathit{Tr}(P_1P^{\prime}_j\ldots,P_S)
  \end{equation}
  \begin{equation}\label{eq:11-2}
    \begin{split}
      {P_j}^{\prime}=P_j \otimes \begin{pmatrix}
        e^{-\beta V_c} & 0\\
        0  & e^{-\beta V_d}
      \end{pmatrix}=\\
      =\begin{pmatrix}
        \exp(-\beta {E^{\prime}}_A) & \exp\left(-\beta(E^{\prime}_A+\mathit{U})\right)\\
        \exp\left(-\beta(E^{\prime}_B+\mathit{U})\right)  & \exp(-\beta {E^{\prime}}_B)
      \end{pmatrix}\mysplit
      \otimes\begin{pmatrix}
        e^{-\beta V_c} & 0\\
        0  & e^{-\beta V_d}
      \end{pmatrix}
    \end{split}
    \end{equation}
\end{subequations}

where  $\otimes$  means outer
product, $V_c$ and $V_d$ are the molecular energies in binding and unbinding state respectively. From Eq\eqref{eq:11-1}, one has 

\begin{equation}\label{eq:12}
Z=(\lambda_{max})^s \left(e^{-\beta V_c}+e^{-\beta V_d}\right)
\end{equation}

instead of Eq\eqref{eq:11-1}. However, the order parameter equations Eq\eqref{eq:12} remain unchanged.  Therefore, the proposed theory can
also be used in analyzing cell fate transitions initiated by small molecules.

\section{Discussion}

Thus all mechanisms of cell fate transitions, be it the reprogramming of fibroblasts into iPSCs, or the activation
of stem cells into proliferative progenitors, or the differentiation of stem cells into terminally differentiated
cells, or the transdifferentiation of cells between different lineages, could be ascribed to quantum transitions
between different DNA macromolecular torsion states. This paradigm could also quantifiably explain why, besides
lineage-specific signaling factors and transcription factors, some apparently non-specific factors like metabolites,
histones and DNA intercalating drugs can also trigger abrupt changes in cell fate programming. A general corollary of
this new paradigm is that a full understanding of all the physiochemical and metabolic factors that control DNA
macromolecular torsion could help us improve our speed and efficiency in directing cell fates for molecular medicine.

One might notice that while we have only discussed two phases of local DNA torsion, $A$ and $B$, there exists a large number
of possible cell fates. This is largely because chromosomal DNA is actually not a homogeneous chain as simply assumed
in this model, but peppered with heterogeneous elements such as insulators, tandem repeat elements (constitutive
heterochromatin), and super-enhancer regions between the gene cluster regions. Hence each insulator, tandem repeat
element, enhancer or gene cluster region could undergo local A-B phase transitions with its own unique torsion
conditions and critical torsion points. This would produce a huge combinatorial diversity of DNA regions in either phase
$A$ or $B$, thus generating a large number of possible cell fates. Others have also noted that tissue development can be
represented as a series of bifurcations into two possibilities on a trajectory in the Waddington ``epigenetic''
landscape\cite{Xu2014}, which is consistent with our two-phase simplification of the DNA torsion theory. 

Recent advances in FTIR imaging, which is a label-free technique being used for detecting specific chemical
compositions via their bonds' vibrational energy spectra and which can now attain nanometer resolutions, should permit
us to repurpose the FTIR technique to observe the DNA torsion energy profile in local regions of a cell's
chromosomes\cite{Ajaezi2018}. If our theory is true, we should then be able to non-invasively observe and predict a cell's epigenetic
state and fate transitions, simply based on just its DNA torsion energy profile. Combined with the latest advances in
partial wave spectroscopic (PWS) microscopy to map local chromatin density profiles\cite{Gladstein2018}, this could have important
implications for medical diagnostics in the future. 

We should also be able to induce abrupt cellular fate transitions with critical concentrations of drugs that
specifically alter the DNA torsional state, either by directly intercalating with DNA and altering its supercoiling, or
by indirectly regulating the histone modifications to alter the supercoiling of DNA around nucleosomes. This could have
important implications for stem cell therapies in regenerative medicine, anti-aging therapies in gerontology, cancer
therapies in oncology, or any therapeutics that involve cellular plasticity. In fact, the DNA intercalating doxorubicin
and other anthracyclines are already well-established anti-cancer drugs, with potent effects on the chromatin state and
DNA supercoiling\cite{Yang2014}. Previous studies indicate that they can treat leukemia, not simply by inducing apoptosis as
previously thought, but by inducing leukemic stem cell differentiation\cite{Larsen1994}. Chromatin-related cancer drugs that inhibit
the HMT/JHDM/HAT/HDAC enzymes, some of which have passed clinical trials, could work in a similar fashion. Recent
findings also suggest that the transition from a proliferative state to aging-associated senescence is due to
reversible defects in chromatin maintenance, with implications for the potential reversal and treatment of aging and
progeria syndromes\cite{Chandra2016}. Even bacterial DNA supercoiling has an impact on bacterial growth and dormancy\cite{Conter2003}, with obvious
implications for our ongoing war with infectious epidemics, and our constant search for new antibiotics and new
therapeutic windows to target multi-drug resistant bacteria. New methods to control cell fate via DNA torsion
transitions could represent a new class of strategies for medical therapeutics in the future. 

With these diagnostic and therapeutic applications in mind, future work could focus on molecular simulations and precise
measurements of DNA torsion potential minima. Such efforts could be based on local chromatin torsion and density
profiles and their associated metabolic (or other physiochemical) variables, in combination with the genomic
information networks they encode, to accurately predict any cellular state and cell fate transition. Given the
generality of DNA supercoiling and cell fate transitions, we expect our DNA torsion-based cellular phase transition
theory to be relevant to almost every field of medicine.

%\bibliographystyle{unsrt}
%\bibliography{References}
%\printbibliography
%

\end{document}